\documentclass[journal]{IEEEtran}
\IEEEoverridecommandlockouts
\usepackage{amssymb}
\usepackage{amsmath}
\usepackage{cite}
\usepackage{siunitx}
\usepackage{comment}
\usepackage{amsfonts}
\usepackage{graphicx}
\usepackage[caption=false,font=footnotesize]{subfig}
\usepackage{textcomp}
\usepackage{xcolor}
\usepackage{float}
\usepackage{makecell}
\usepackage{booktabs}

\def\BibTeX{{\rm B\kern-.05em{\sc i\kern-.025em b}\kern-.08em
    T\kern-.1667em\lower.7ex\hbox{E}\kern-.125emX}}
\begin{document}

\title{A Survey on Robust Deep Joint Source-Channel Coding for Semantic Communications}

\author{
		\IEEEauthorblockN{Eunhye Hong, Taewoo Park, and Yongjune Kim}	\\	    
    \thanks{                
        Eunhye Hong, Taewoo Park, and Yongjune Kim are with the Department of Electrical Engineering, Pohang University of Science and Technology (POSTECH), Pohang 37673, South Korea (e-mail: \{eunhye.hong, parktaewoo, yongjune\}@postech.ac.kr). 
    }  
}

\maketitle

\begin{abstract}
Semantic communications (SCs) aim to transmit only the essential information required to perform given tasks, thereby improving communication efficiency. 
Deep learning-based joint source–channel coding (deep JSCC) has emerged as a promising approach for SC systems; however, its performance often degrades when the deployment channels differ from the training channel conditions, making robustness a critical requirement.
This paper presents a structured overview of recent methodologies for enhancing the robustness of deep JSCC. 
Specifically, existing approaches are categorized into two classes: robust training approaches and adaptive approaches, with the latter further divided into adaptive semantic feature selection, physical-layer adaptation, and semantic feature adaptation.
Finally, we discuss promising directions, including multi-task generalization and explainability in robust SC systems.
\end{abstract}


\maketitle

\section{Introduction}

Recent advances in artificial intelligence (AI) have driven the development of applications such as autonomous driving and augmented reality (AR), which require ultra-reliable and low-latency communications (URLLC)~\cite{Saad2020vision, Hong20226g}. 
Conventional communication systems, grounded in the Shannon paradigm, are designed to ensure accurate bit-level transmission. 
However, this focus on bit-level fidelity often leads to excessive redundancy and latency, thereby becoming a bottleneck for AI-driven applications.
To address these limitations, a paradigm shift toward semantic communications (SCs) has been proposed, focusing on transmitting task-relevant semantic information~\cite{Gunduz2022beyond, Shi2023task}.
Rather than converting all source data into bits and treating them equally during transmission, SCs selectively extract and convey the information most relevant to the task objective.
This enables a substantial reduction in communication overhead while still achieving high task performance.

SCs can be broadly categorized into three types: data recovery, task execution, and joint data recovery and task execution~\cite{You2025next}.
Data recovery aims to reconstruct transmitted data with high semantic accuracy~\cite{Bourtsoulatze2019deep, Choi2019neural, Kurka2020deep}; task execution focuses on performing a specific task at the receiver without necessarily recovering the entire data~\cite{Shao2022learning, Xie2023robust, Hu2023robust, Kim2023distributed}; and the joint approach seeks to support both data reconstruction and task execution, representing a trade-off between accurate reconstruction and task performance~\cite{huang2024joint, Lyu2024semantic}.

\begin{figure*}[t!]
    \centering
    \includegraphics[width=0.75\textwidth]{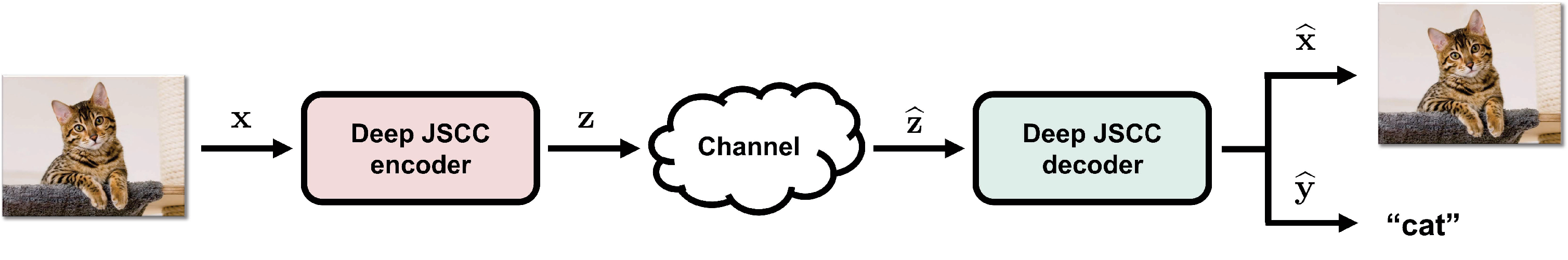}
      \caption{System model of a deep JSCC framework for semantic communications. }
    \label{fig:system_model}
\end{figure*}

An important research topic in this paradigm is deep joint source-channel coding (deep JSCC), which integrates compression and error protection within neural networks~\cite{Bourtsoulatze2019deep, Shao2022learning, Xie2021deep}.
In deep JSCC, the encoder maps input data directly to channel symbols, and the channel is modeled as a non-trainable layer in the network.
The decoder then produces the required output for the downstream task from the noisy received semantic feature, and the entire system is jointly optimized in an end-to-end manner.
Unlike traditional separate source and channel coding, deep JSCC employs an end-to-end design that mitigates the cliff effect by ensuring graceful performance degradation~\cite{Bourtsoulatze2019deep}.
Building on these advances, deep JSCC schemes have been developed for various data modalities, including image~\cite{Bourtsoulatze2019deep, Choi2019neural, Kurka2020deep, Shao2022learning, Xie2023robust, Xu2022wireless}, text~\cite{Farsad2018deep, Xie2021deep}, audio~\cite{Weng2021semantic}, and multi-modal data~\cite{Zhang2024unified, He2024rate} transmission.

However, the very end-to-end nature that underpins the advantages of deep JSCC also introduces a fundamental limitation: a strong dependence on the channel statistics assumed during training~\cite{Bourtsoulatze2019deep, Shao2022learning, Xie2023robust, Xu2022wireless}.
Consequently, a mismatch between the training and deployment channel conditions--commonly referred to as the \emph{channel mismatch} problem--can cause severe degradation in task performance.
Since channel conditions in practice vary unpredictably over time and space, this issue becomes even more pronounced.
Deploying separate models for every possible channel is clearly impractical, making it imperative to develop SC systems that remain robust under diverse and dynamic channel variations.

Given the unpredictable nature of wireless channels, achieving robustness to channel variations is critical for reliable SCs.
Motivated by this need, this paper provides a structured overview of recent methodologies for robust SC systems.
While existing survey articles on SCs mainly provide broad overviews of SC frameworks, applications, and enabling techniques, this survey focuses specifically on methodologies for improving robustness to channel mismatch and channel variation in deep JSCC based SCs.
We first describe the system model of SCs to establish a common basis for discussion. We then categorize representative studies into two main groups. This categorization is based on whether the trained model and transmission strategies remain fixed after training or are allowed to vary with channel conditions during operation:
\begin{itemize}
\item \textbf{Robust training approaches}, which improve model robustness by learning encoder-decoder mappings that remain reliable across channel variations through training strategies such as loss design and adversarial training.
\item \textbf{Channel-aware adaptive approaches}, which improve robustness by enabling the system to adapt its transmission process or semantic feature processing to channel variations, typically based on available channel information and, in some cases, semantic importance.
This category can be further divided into three subtypes:
(a) \textit{semantic feature selection}, which regulates the number or dimension of transmitted semantic features according to channel conditions; 
(b) \textit{physical-layer adaptation}, which adjusts physical-layer parameters such as bandwidth, power, and modulation order based on channel conditions and semantic importance; and
(c) \textit{semantic feature adaptation}, which refines the semantic features using information about the channel condition.
\end{itemize}

This survey concludes with a discussion of open challenges and future research directions, including the interplay among robustness, multi-task generalization, and explainability.

\section{System Model}

Throughout this paper, we adopt the following notation. 
The source vector is denoted by
$X^n = (X_1, \ldots, X_n)$, and $\mathbf{x}$ represents a realization of $X^n$. 
Similarly, the latent semantic feature vectors $Z^k$ and $\hat{Z}^k$ denote random variables corresponding to the transmitted and corrupted semantic features, with their realizations written as $\mathbf{z}$ and $\hat{\mathbf{z}}$, respectively.

Deep JSCC provides an end-to-end mapping that generates a transmitted semantic feature at the encoder and produces the desired output at the receiver (see Figure~\ref{fig:system_model}).
The semantic encoder, parameterized by $\phi$, maps the $n$-dimensional input data $\mathbf{x}$ to a $k$-dimensional semantic feature vector $\mathbf{z}$.
This encoding process extracts the information relevant to the system objective and represents it in a compact semantic feature vector, typically with $k \ll n$, thereby enabling efficient bandwidth usage.
The encoded feature is then transmitted over a channel characterized by the conditional transition distribution $p_{\text{ch}}(\hat{\mathbf{z}} \mid \mathbf{z})$.

The decoder depends on the system objective, and its detailed form is described in the following subsections.

\subsection{Data Recovery}

For data recovery, the end-to-end information flow can be modeled as 
\begin{equation}
    X^n \rightarrow Z^k \rightarrow \hat{Z}^k \rightarrow \hat{X}^n,
    \label{eq:recovery_chain}
\end{equation}
where $X^n \sim p(\mathbf{x})$.
The transmitter observes the source data $\mathbf{x}$ and encodes it into a semantic feature vector $\mathbf{z}=f_\phi(\mathbf{x})$ that preserves information useful for source reconstruction.
The receiver takes the corrupted feature $\hat{\mathbf{z}}$ as input and produces the reconstructed data $\hat{\mathbf{x}}=g_\theta(\hat{\mathbf{z}})$.
The training objective is to minimize the reconstruction error:
\begin{equation}
    \mathcal{L}_{\text{rec}}(\phi,\theta) = \mathbb{E}_{p(\mathbf{x})p_{\text{ch}}(\hat{\mathbf{z}}|\mathbf{z})}\left[\ell_{\text{rec}}\left(\mathbf{x},\hat{\mathbf{x}}\right)\right],
\end{equation}
where $\ell_{\text{rec}}\left(\mathbf{x},\hat{\mathbf{x}}\right)$ measures the distortion between the source and the reconstruction, typically employing metrics such as mean squared error (MSE) for pixel-level fidelity or structural similarity (SSIM) for perceptual quality.

\subsection{Task Execution}

For task execution, deep JSCC directly performs the target task without reconstructing the entire data. 
The end-to-end process is modeled as
\begin{equation}
    Y \rightarrow X^n \rightarrow Z^k \rightarrow \hat{Z}^k \rightarrow \hat{Y},
    \label{eq:task_chain}
\end{equation}
where $(X^n, Y) \sim p(\mathbf{x},{y})$ denotes the source vector $X^n$ paired with its associated task label $Y$.
In practice, the transmitter does not know the task label $y$; instead, it observes the source data $\mathbf{x}$ and encodes it into a task-relevant semantic feature vector $\mathbf{z}=f_\phi(\mathbf{x})$.
After obtaining the corrupted semantic feature $\hat{\mathbf{z}}$, the receiver predicts the task output $\hat{y} = g_\theta(\hat{\mathbf{z}})$.
The training objective is
\begin{equation}
    \mathcal{L}_{\text{task}}(\phi,\theta) = \mathbb{E}_{p(\mathbf{x},y)p_{\text{ch}}(\hat{\mathbf{z}}|\mathbf{z})}\left[\ell_{\text{task}}\left(y,\hat{y}\right)\right],
\end{equation}
where $\ell_{\text{task}}$ denotes a loss function tailored to the downstream task, such as cross-entropy for classification.

\subsection{Joint Data Recovery and Task Execution}

For the joint objective, the process is modeled as
\begin{equation}
    Y \rightarrow X^n \rightarrow Z^k \rightarrow \hat{Z}^k \rightarrow \left(\hat{X}^n, \hat{Y}\right),
\end{equation}
where $(X^n, Y) \sim p(\mathbf{x}, y)$. 
The transmitter encodes the source data $\mathbf{x}$ into a shared semantic feature vector $\mathbf{z}=f_\phi(\mathbf{x})$ that supports both accurate source reconstruction and reliable task execution at the receiver.
Based on the shared corrupted feature vector $\hat{\mathbf{z}}$, the receiver reconstructs the source data $\hat{\mathbf{x}}$ and predicts the task output $\hat{y}$, potentially using separate decoders for each objective, whose parameters are collectively denoted by $\theta$.
The combined training objective is formulated as
\begin{equation}
    \mathcal{L}_{\text{joint}}(\phi,\theta) =
    \lambda \mathcal{L}_{\text{rec}}(\phi,\theta)+(1-\lambda)\mathcal{L}_{\text{task}}(\phi,\theta),
    \label{eq:joint_loss}
\end{equation}
where $\lambda\in[0,1]$ balances the trade-off between reconstruction fidelity and task performance.

\section{Channel Mismatch Problem in Deep JSCC}

In deep JSCC, the channel is incorporated into the end-to-end learning pipeline as a non-trainable layer.
Accordingly, the encoder and decoder are optimized with respect to the channel statistics assumed during training, and the learned mapping tends to become specialized to the channel statistics encountered during training.
However, in realistic deployment scenarios, wireless channels fluctuate over time and space due to mobility, fading, interference, and hardware impairments. 
Consequently, the actual deployment channel often deviates significantly from the training channel model. 
When such a discrepancy occurs, the resulting \emph{channel mismatch} can cause severe degradation in both reconstruction fidelity and task performance.
Deploying separate models for every possible channel is clearly impractical, posing challenges in accommodating diverse and dynamic channel variations.

A commonly used approach to mitigate this issue is to train the model on data generated from a mixture of diverse channel conditions, enabling it to encounter a broader range of perturbations during learning~\cite{Bourtsoulatze2019deep}. 
Although such mixture training can partially improve robustness, it generally underperforms models optimized for the actual deployment channel and does not fully resolve the mismatch problem, especially when the channel varies dynamically after deployment.
These limitations motivate the need for mechanisms that enhance robustness to channel variations while maintaining reliable performance, which is essential for practical deep JSCC systems.

\section{Methods for Enhancing Robustness to Channel Variations}

To enhance robustness against channel mismatch, several approaches have been proposed. 
In this survey, we provide a structured overview of these methodologies by categorizing them into two classes: robust training approaches and adaptive approaches. 
The distinction is based on whether the trained model and transmission rule remain fixed after training or whether part of the transmission process or semantic feature processing varies with channel conditions during operation.
Robust training approaches improve robustness by learning encoder-decoder mappings that remain reliable across channel variations while keeping the trained model and transmission rule fixed after training. 
Adaptive approaches improve robustness by enabling the system to adjust the number of transmitted semantic features, physical-layer transmission parameters, or the semantic feature representation according to the current channel condition during inference.
Table~\ref{tab:robust_summary} summarizes the proposed categorization and indicates whether each approach requires channel information during inference.

\begin{table*}[ht!]
    \centering
    \small
    \caption{Categorization of Robust Semantic Communication Schemes Against Channel Variations}
    \label{tab:robust_summary}
    \renewcommand{\arraystretch}{1.15}
    \begin{tabular}{
        >{\centering\arraybackslash}m{4cm}
        >{\centering\arraybackslash}m{5cm}
        >{\centering\arraybackslash}m{3cm}
        >{\centering\arraybackslash}m{3cm}}
        \toprule
        Category & Channel Information & Data Recovery & Task Execution\\
        \toprule

        Robust Training Approaches & 
        Not Required & 
        --&\cite{Xie2023robust, Park2025regularized, Park2025robust, Wei2025robust} \\ 
        \midrule

        Semantic Feature Selection & 
        Required & \cite{Yang2022deep,  Zhang2023predictive} &
        \cite{Shao2022learning} \\ 
        \midrule

        Physical-Layer Adaptation & 
        Required & \cite{Wu2024deep} &
        \cite{Gao2024rate, Park2025joint} \\ 
        \midrule

        Semantic Feature Adaptation & 
        Required & \cite{Ding2021snr, Xu2022wireless, Xie2024deep, Zhang2025snr} &
        \cite{Xie2024deep} \\

        \bottomrule
    \end{tabular}
\end{table*}

\subsection{Robust Training Approaches}

Robust training approaches aim to improve the robustness of deep JSCC models under varying channel conditions.
The central idea is to learn an encoder-decoder mapping that the model can sustain reliable performance when the channel conditions encountered during deployment differ from those assumed in training.
To this end, a range of training strategies, including loss function design~\cite{Xie2023robust,Park2025robust, Park2025regularized} and adversarial training~\cite{Wei2025robust}, have been proposed to reduce sensitivity to channel mismatch and improve robustness.

One line of work improves robustness by leveraging coded redundancy to utilize channel capacity more effectively. 
Xie et al.~\cite{Xie2023robust} introduced the robust information bottleneck (RIB) framework, which enhances robustness by embedding coded redundancy into semantic features. 
This framework is implemented through discrete task-execution JSCC (DT-JSCC) using vector quantization compatible with digital modulation.
The RIB principle regulates the trade-off between task performance and model robustness through a tunable hyperparameter that balances between task-relevant information and coded redundancy.

Another line of research addresses robustness by stabilizing the posterior distribution that governs inference under channel perturbations.
Channel variations can distort the posterior distribution conditioned on the transmitted semantic feature, leading to degraded task performance.
Motivated by this observation, Park et al.~\cite{Park2025robust, Park2025regularized} proposed a principled approach that improves robustness by explicitly minimizing the discrepancy between the posterior distributions computed from noise-free and noisy semantic features.
This discrepancy is measured using the Kullback-Leibler (KL) divergence, which is incorporated into the training objective to enforce prediction consistency under channel perturbations.
The resulting KL divergence is further approximated via a second-order Taylor expansion, yielding a form proportional to the channel noise variance and the trace of the Fisher information matrix.
Such a formulation effectively smooths the log-posterior, leading to consistent predictions even when the semantic feature is distorted by channel noise.
Moreover, the contribution of this KL term is adaptively scaled with the channel noise variance, enabling the model to adjust its robustness according to the channel condition.

To enhance robustness under diverse channel conditions, Wei et al.~\cite{Wei2025robust} adopted adversarial training so that the semantic encoder-decoder to learn semantic features that are resilient to a wide range of channel distortions encountered during training.
As a result, the model can improve generalization across different channel environments without relying on deeper architectures or explicit adaptation modules.

\subsection{Channel-Aware Adaptive Approaches}

Channel-aware adaptive approaches enhance robustness by dynamically adjusting part of the transmission process or semantic feature processing under varying channel conditions, typically using available channel information and, in some cases, semantic importance.
These methods adapt the transmission behavior at runtime to maintain semantic fidelity and task performance under time-varying or uncertain environments.
Based on the level at which adaptation is applied, existing adaptive approaches can be categorized into three types:
(i) \textit{semantic feature selection}, which controls the number or dimension of transmitted semantic features;
(ii) \textit{physical-layer adaptation}, which adjusts transmission parameters at the physical layer, such as power allocation, modulation order, or subchannel usage, based on channel conditions and semantic importance; and
(iii) \textit{semantic feature adaptation}, which modifies the internal processing or representation of semantic features based on information about the channel condition.
These adaptive mechanisms enable deep JSCC systems to sustain reliable performance across diverse channel environments.

\subsubsection{Semantic Feature Selection}
% compression ratio control
Adaptive semantic feature selection approaches enhance robustness by dynamically regulating the number or dimensionality of transmitted semantic features according to channel conditions and communication constraints.
When the channel quality deteriorates, the encoder transmits more features to provide a richer description of the underlying semantics. 
This allows the decoder to recover semantic information even when part of the transmitted features are distorted by noise, resulting in more stable task performance under channel variation. 
Conversely, when the channel condition is favorable, fewer features can be transmitted since less distortion occurs during transmission, and the semantic information can still be reliably recovered from a more compact feature.
By adaptively adjusting the number of transmitted semantic features, the system maintains robustness while improving communication efficiency.

Shao et al.~\cite{Shao2022learning} proposed the variable-length variational feature encoding (VL-VFE) framework as an adaptive feature selection method. 
VL-VFE extends the original variational feature encoding (VFE), which is grounded in the information bottleneck principle and employs a sparse prior to represent task-relevant information with a compact set of latent features.
By dynamically adjusting the number of activated neurons in response to channel conditions, VL-VFE enables adaptive transmission without modifying the model architecture.
This adaptive feature selection mechanism allows VL-VFE to maintain task performance across varying channel conditions.

Yang and Kim~\cite{Yang2022deep} proposed an adaptive deep JSCC framework in which the number of transmitted feature groups is determined jointly by the channel SNR and the semantic complexity of the input image.
The encoder partitions the semantic feature into non-selective and selective feature groups.
Non-selective features are always transmitted to preserve essential semantic content, whereas selective features are transmitted only when necessary.
A policy network determines the number of selective feature groups to transmit based on the observed SNR and the content-dependent importance of the image.
This design enables adaptive control of the effective compression ratio, allowing the system to transmit more features under challenging channel conditions or semantically complex input, and fewer features when the channel is reliable or the input image is semantically simple.

Selecting semantic features to satisfy a target reconstruction quality is challenging in deep JSCC, as the reconstruction quality is jointly determined by the channel SNR, the compression ratio, and the semantic content of the data.
To address this challenge, Zhang et al.~\cite{Zhang2023predictive} proposed predictive and adaptive deep coding (PADC), a framework for quality-aware feature selection in deep JSCC-based image transmission. 
PADC integrates a variable-length DeepJSCC-V model with an oracle network (OraNet) that predicts the reconstruction PSNR based on image content, channel SNR, and compression ratio.
Using these predictions, a compression-ratio optimizer selects the minimum compression ratio that satisfies a target PSNR requirement. 
This predictive mechanism enables feature selection for each image and allows the system to maintain the desired reconstruction quality across diverse channel conditions and image contents.

\subsubsection{Physical-Layer Adaptation}

In SC systems, physical-layer resources such as bandwidth transmit power, and modulation order directly affect the available transmission capacity and affect the reliability of wireless communications.
Adaptive management of physical-layer resources enables the system to sustain performance under varying channel conditions.
In this approach, robustness is improved by adapting physical-layer transmission parameters or resource usage according to channel conditions and semantic requirements of the transmitted information.

In~\cite{Gao2024rate,Park2025joint}, modulation adaptation schemes are proposed that select the transmission rate based on the robustness of the downstream semantic model.
These methods determine the maximum perturbation of the received signal that the model can tolerate.
Given this tolerance, the system evaluates the bit error rate (BER) for each modulation order under the current SNR and selects the highest order whose BER remains within the verified tolerance range.
When the model is tolerant to larger distortions, a higher modulation order is chosen to improve efficiency, whereas a more sensitive model prompts selection of a lower modulation order to maintain task performance.

Wu et al.~\cite{Wu2024deep} proposed a framework that jointly integrates semantic feature learning with physical-layer channel awareness.
The model augments a Vision Transformer encoder with a channel heatmap that characterizes the reliability of multi-input multi-output (MIMO) subchannels.
This heatmap is embedded jointly with the image tokens and processed through the Transformer's self-attention layers, enabling the encoder to shape the semantic feature such that semantically important components are naturally aligned with more reliable spatial-frequency paths of the MIMO channel.
By explicitly incorporating channel state information, the encoder-decoder pair remains effective across varying SNR levels, antenna configurations, and channel estimation accuracy without requiring retraining.

\subsubsection{Semantic Feature Adaptation}

Channel-aware semantic feature adaptation improves robustness by modifying the internal processing or representation of semantic features using information about the channel condition.
In these methods, such channel information is mainly provided in the form of the estimated SNR and is incorporated during feature extraction, refinement, or decoding, enabling the processing semantic features to adapt to the current channel environment.
As a result, the feature-processing behavior varies with SNR, allowing a single deep JSCC model to operate across different SNR levels without modifying the transmission rate or the model architecture.

Ding et al.~\cite{Ding2021snr} proposed an SNR-adaptive deep JSCC framework in which the decoder estimates the SNR from pilot signals and uses this estimate to construct an SNR map that is combined with the received semantic features.
The decoder incorporates a denoising module that exploits the SNR information, allowing the denoising strength to vary with the noise level.
By adapting the denoising operation according to the estimated SNR, the proposed method improves reconstruction quality across a wide range of channel conditions.

Subsequent work explored how both the encoder and decoder can adjust their feature processing according to the channel quality.
Xu et al.~\cite{Xu2022wireless} proposed an attention deep JSCC (ADJSCC) framework in which the encoded features are adjusted according to the channel condition.
The ADJSCC model interleaves feature learning (FL) modules with attention feature (AF) modules.
Each AF module extracts global feature statistics, combines them with the estimated SNR, and generates scaling values that adjust the strength of the feature channels.
Through this design, the encoder can handle different channel qualities by applying adjustments that depend on the estimated SNR during the encoding process.

Recent work considers adaptation strategies in which the encoder and decoder parameters change with the channel condition.
Xie et al.~\cite{Xie2024deep} proposed a hypernetwork-based deep JSCC framework in which the encoder and decoder parameters are generated according to the channel SNR. 
Instead of training separate models for different channel qualities, the hypernetwork produces the parameter values that the backbone network uses at each SNR level. 
This enables the model to adjust its encoding and decoding operations in real time without requiring retraining, and can be applied to both data recovery systems and task execution systems.

Another direction integrates SNR information into the multi-head attention mechanism of Transformer models so that the attention operation changes with the channel quality.
Zhang and Tao~\cite{Zhang2025snr} proposed a query-based deep JSCC framework that incorporates the estimated SNR into the multi-head attention mechanism. 
The estimated SNR is embedded into the attention inputs and is also used to generate queries conditioned on the SNR, allowing the attention computation to reflect the channel condition without modifying the network structure.
In low SNR environments, the model needs to suppress less important elements more strongly to preserve the semantic components required for the task.
To capture this behavior, the framework makes the attention score adjustments depend on the SNR, and the loss function includes penalty terms to stabilize this behavior during training.

\section{Discussion and Future Research Directions}

The studies reviewed in this paper highlight recent progress in robust deep JSCC-based SCs, while also revealing remaining challenges.
In this section, we discuss several open challenges and promising research directions that are expected to shape the future of SCs.

\subsection{Multi-Task Generalization}

Most existing robust SC systems are optimized for a single, predefined task.
In practice, however, transmitted semantic information often needs to support multiple downstream tasks simultaneously.
To reduce communication overhead, one possible approach in multi-task SC systems is to rely on a  shared encoded semantic feature that is reused across tasks.
This becomes challenging because different tasks may require different kinds of semantic information.
This tendency is particularly evident in computer vision task.
For example, image classification mainly requires  information about what is present in image, whereas image retrieval requires a broader understanding of the overall image content~\cite{Park2025transmit}.
This observations suggest that, when multiple tasks share a common semantic representation, different tasks may rely on different components of that representation.
Accordingly, channel perturbations affecting particular semantic components may not affect all tasks equally.
Tasks that rely more heavily on the corrupted components may suffer more severe performance degradation than others.
This uneven sensitivity to channel noise across tasks makes robustness particularly challenging in multi-task SC systems.
From this perspective, an important direction for future research is to design SC systems that ensure robustness across multiple tasks over noisy channels.

\subsection{Explainability and Robustness}

Most SC systems operate largely as black boxes, with learned semantic features that are difficult to interpret.
This lack of explainability has motivated recent efforts to introduce explainability into SC models, aiming to better understand and control the semantic features being transmitted.
Ma et al.~\cite{Ma2023task} addressed this limitation by proposing an explainable framework that learns disentangled semantic feature using a $\beta$-VAE, thereby improving the interpretability of semantic features.
Im et al.~\cite{Im2024attention} leveraged attention scores from Vision Transformers to identify image patches that are most relevant to the classification task.
By selectively transmitting these salient patches, the communication cost can be reduced while preserving inference accuracy.

Explainability has the potential to play a critical role in robust SC systems by enabling the identification of semantically important information.
In particular, selectively protecting such critical and interpretable features can improve both communication efficiency and robustness.
Building on this perspective, an open research direction is to develop principled approaches that leverage explainability to improve robustness in SC systems.

\section{Conclusion}

This paper presented a survey of robust SCs under channel mismatch and variability. 
The growing integration of AI-driven applications with wireless networks requires SC systems that sustain reliable performance under unpredictable channel conditions without incurring excessive redundancy or latency. 
The reviewed studies collectively indicate that a single model optimized for a fixed operating point is insufficient for practical deployment.

We categorized existing methods into two main classes.
First, robust training approaches aim to improve model robustness through techniques such as loss function  design and adversarial training, thereby reducing the model's sensitivity to channel perturbations.
Second, channel-aware adaptive approaches dynamically adjust the semantic encoding and transmission process at runtime by leveraging side information such as SNR, CSI, or semantic importance.
This class includes semantic feature selection, physical-layer adaptation schemes, and semantic feature adaptation.

We also identified critical open challenges for future research, including multi-task generalization and the potential synergy between explainability and robustness.
The development of these robust SC systems would represent an important step toward realizing intelligent and efficient 6G networks, in which communication is seamlessly integrated with the execution of AI tasks.

\bibliographystyle{IEEEtran}
\bibliography{abrv,mybib}
\end{document}